\documentclass[twocolumn,showpacs,aps,prd,superscriptaddress,amsmath,amssymb,floatfix]{revtex4}
\usepackage{graphicx}
\usepackage{bm}
\begin{document}
\title{Contribution of low-lying vector resonances to polarization
observables in ${\bar B}_d^0\to {\bar K}^{*0}\,e^+\,e^-$ decay}
\author{Alexander Yu. Korchin} \email{korchin@kipt.kharkov.ua}
\affiliation{NSC `Kharkov Institute of Physics and Technology',
61108 Kharkov, Ukraine}
\author{Vladimir A. Kovalchuk}  \email{koval@kipt.kharkov.ua}
\affiliation{NSC `Kharkov Institute of Physics and Technology',
61108 Kharkov, Ukraine}
\date{today}

\begin{abstract}
The branching ratio and other observables for the rare
flavor-changing neutral current decay ${\bar B}_d^0 \to {\bar
K}^{*0} \, (\to K^{-}\, \pi^+) \, e^+\,e^- $  are studied below
the $\bar{c} c$ threshold. The total amplitude for this decay
includes the term coming from the standard model effective
Hamiltonian and the term generated by the processes $\bar{B}_d^0
\to \bar{K}^{*0} \, (\to K^{-}\, \pi^+) \, V $ with intermediate
low-lying vector resonances $V = \rho(770), \, \omega(782), \,
\phi(1020)$ decaying into the $e^+ e^-$ pair. The resonance
contribution to the branching ratio, polarization fractions of the
$K^*$ meson, and coefficients in the angular distribution is
calculated. The influence of the resonances on the integrated
observables in the region of electron-positron invariant mass up
to 1 GeV is studied in view of the planned measurements of the
photon polarization at the LHCb.
\end{abstract}


\maketitle

\setcounter{footnote}{0}

\section{\label{sec:Introduction}Introduction}
The investigation of rare $B$ decays induced by the
flavor-changing neutral current (FCNC) transitions $b\to s$ and
$b\to d$ represents an important test of the standard model (SM)
and its extensions (see \cite{Antonelli:2009} for a review). Among
the rare decays, the radiative decay $b\to s\gamma$ has probably
been the most popular FCNC transition ever since its experimental
observation as $B\to K^*\gamma$ at CLEO in 1993 \cite{CLEO:1993}.
This decay proceeds through a loop (penguin) diagram, to which
high-mass particles introduced in extensions to the SM may
contribute with a sizable amplitude. The size of the decay rate
itself, however, provides only a mild constraint on such
extensions, because the SM predictions for exclusive rates suffer
from large and model-dependent form factor uncertainties
\cite{Ali:2002, Bosch:2002}. Further reduction in the errors of
the theory appears rather difficult. It is then clearly
advantageous to use, in addition to the rates, other observables
that can reveal new physics (NP).

In particular, in the framework of the SM, the photons emitted in
$b\to s\gamma$ decays are predominantly left-handed, while those
emitted in $\bar b$ decays are predominantly right-handed. Based
on the leading order effective Hamiltonian, the amplitude for
emission of wrong-helicity photons is suppressed by a factor
$\propto m_s/m_b$ \cite{Atwood:1997}. This suppression can easily
be alleviated in a large number of NP scenarios where the helicity
flip occurs on an internal line. An independent measurement of the
photon helicity is therefore of interest. Several different
methods of measuring the photon polarization have been suggested.
In one method the photon helicity is probed through mixing-induced
CP asymmetries \cite{Atwood:1997}. Another method makes use of the
photons from the $B\to \gamma\,K^*(\to K\pi)$ decay, which are
converted into the electron-positron pair in the detector material
\cite{Grossman:2000, Seghal:2004}. There are also other techniques
to probe photon polarization. These include approaches in which
interference between different resonances \cite{Gronau:2002} or
different helicity states \cite{Atwood:2007} of the hadronic
recoil system provide sensitivity to the polarization. The photon
polarization may also be studied in radiative decays of
$\Lambda_b$ baryons \cite{Mannel:1997}. It appears, however, that
experimentally the photon polarization is difficult to measure,
and one instead has to use the process $b\to s \gamma^* \to
s\ell^+\ell^-$, where the photon is converted to the lepton pair.
In this decay the angular distributions and lepton polarizations
can probe the chiral structure of the matrix element
\cite{Grossman:2000, Melikhov:1998, Ali:2000, Kruger:2005,
Bobeth:2008, Altmannshofer:2009, Egede:2010} and thereby the NP
effects.

In order to unambiguously measure effects of NP in the process
$b\to s\ell^+\ell^-$, if they indeed show up in the observables,
one needs to calculate the SM predictions with a rather good
accuracy. In general, the SM amplitude consists of the
short-distance (SD) contributions and the long-distance (LD) ones.
The former are expressed in terms of the Wilson coefficients $C_i$
calculated in perturbative QCD up to a certain order in
$\alpha_s(\mu)$; they carry information on processes at energy
scales $ \sim m_W, \ m_t$ [here $\alpha_s(\mu)$ is the effective
QCD coupling constant]. These coefficients are then evolved, using
the renormalization group methods, to the energies related to the
bottom quark mass $m_b$.

The LD terms include factorizable and nonfactorizable effects from
virtual photons via the semileptonic operators ${\cal O}_{9{\rm
V},\,10{\rm A}}$ and electromagnetic dipole penguin operator
${\cal O}_{7\gamma}$ in the effective Hamiltonian. The radiative
corrections coming from the operators ${\cal O}_{1-6}$ and the
gluon penguin operator ${\cal O}_{8g}$ are also accurately
accounted for (for a review, see \cite{Buchalla:1996}).

The LD effects describing the hadronization process are expressed
in terms of hadronic matrix elements of the $b \to s$ operators
between the initial $B$ and the $K^*$ final state. These matrix
elements are parametrized in terms of form factors \cite{Ali:2000}
that are calculated with the help of light-cone sum rules (LCSR)
\cite{Ball:2005} or in soft-collinear effective theory
\cite{Defazio:2006}. The form factors have large theoretical
uncertainties that are presently the dominant uncertainties in the
SM predictions for exclusive decays.

The presence of additional LD effects originating from
intermediate vector resonances $\rho (770)$, $\omega(782)$,
$\phi(1020)$, $J/\psi(1S)$, $\psi(2S)$,$\ldots$ complicates the
description and makes it more model dependent. These resonances
show up in the region of relatively small dilepton invariant mass
$m_{ee} \equiv \sqrt{q^2}$, where $q^2=(q_++q_-)^2$. In order to
suppress the charmonia contribution, often the region of large
dilepton mass ($q^2 \gg 4 m_c^2 \approx 6.5$ GeV$^2$) is selected;
for example, BaBar and Belle Collaborations apply the
corresponding experimental cuts~\cite{Babar:2009, Belle:2009}. In
some cases the resonances $J/\psi (1S), \ \psi (2S)$ are
explicitly excluded in the analysis via the Breit-Wigner energy
factors.

The region of small dilepton invariant mass, $m_{ee} \lesssim 1$
GeV, has attracted less attention so far. Nevertheless, as was
pointed out in \cite{Grossman:2000}, this region also has a high
potential for searching for NP effects. At small $m_{ee} \sim M_R$
the low-lying vector resonances modify the amplitude and thus may
induce, in certain observables, the right-handed photon
polarization, which is still small but not negligible. The
presence of the photon propagator $1/q^2$ enhances the resonance
contribution. Recently, the authors of \cite{Lefrancois:2009}
analyzed the angular distribution in the rare decay $\bar{B}^0 \to
\bar{K}^{*0} e^+ e^-$ in the small-$q^2$ region, in order to test
the possibility to measure this distribution at the LHCb. They
have shown the feasibility of future measurements with small
systematic uncertainties.

In the present paper we calculate the branching fraction $d \Gamma
/dq^2$ and asymmetries in the $\bar{B}^0 \to \bar{K}^{*0} e^+ e^-$
decay at dilepton invariant mass $m_{ee} < 2.5$ GeV. Both the SD
and LD effects in the amplitude are evaluated. We use the
effective Hamiltonian with the Wilson coefficients in the
next-to-next-to-leading order (NNLO) approximation. The LD effects
mediated by the resonances, {\it i.e.} $\bar{B}^0 \to \bar{K}^{*0}
V \to \bar{K}^{*0} \gamma^* \to \bar{K}^{*0} e^+ e^-$ with
$V=\rho(770), \ \omega(782), \ \phi(1020)$, are included
explicitly in terms of amplitudes of the decays $\bar{B}^0 \to
\bar{K}^{*0} V $. The information on the latter amplitudes is
taken from experiment if available; otherwise it is taken from
theoretical predictions.

We also study the sensitivity of the observables in the $\bar{B}^0
\to \bar{K}^{*0} e^+ e^-$ decay to the choice of the form factors
of the transition $B \to K^*$. In the literature there exists a
large variety of models for these form factors. We choose a few
models \cite{Ball:2005, Ali:2000, Charles:1999, Beneke:2001} in
our calculation. The other nontrivial aspect of the theory is the
mass of the strange quark $m_s$, as a nonzero value of $m_s$ leads
to a small admixture of the right-handed photon polarization.
Therefore, we calculate observables with both zero and nonzero
values of the strange quark mass.

We calculate the coefficients $A_{\rm T}^{(2)}$ and $A_{\rm Im}$,
which determine, respectively, $\cos (2 \phi)$ and $\sin (2 \phi)$
dependencies in the angular distributions of the leptons ($\phi$
is the angle between the plane spanned by $e^+, \ e^-$ and the
plane spanned by the decay products $K^-, \ \pi^+$ of the
$\bar{K}^{*0}$ meson). The other observables, such as
forward-backward asymmetry $d A_{\rm FB}/dq^2$ and polarization
parameters of $K^*$ meson $f_{0}, \ f_{\|}, \ f_{\perp}$, are also
calculated.

The paper is organized as follows. In Sec.~\ref{sec:formalism} the
main formulas for the calculation of observables are presented. In
Sec.~\ref{subsec:angle distribution} the expressions for the fully
differential decay rate and partially integrated ones over the
angles and the dilepton invariant mass are given.
Section~\ref{subsec:transversity} contains expressions for
transversity amplitudes in the SM, and the amplitudes in the limit
of very small $q^2$. Contributions to the amplitudes from
resonances $\rho(770), \ \omega(782), \ \phi(1020)$ and all
ingredients needed for their calculation, are discussed in
Sec.~\ref{subsec:resonances}. Results of the calculations and a
discussion are presented in Sec.~\ref{sec:results}. In
Sec.~\ref{sec:conclusions} we draw our conclusions. In
Appendix~\ref{sec:Appendix} some details of the calculation of the
matrix element and the models of the $B \to K^*$ transition form
factors are described.

\section{\label{sec:formalism} Angular distributions and
amplitudes for the ${\bar B}_d^0\to {\bar K}^{*0}\,e^+\,e^-$ decay
}

\subsection{ \label{subsec:angle distribution}
Differential decay rate}

\begin{figure}
\centerline{\includegraphics[width=.45\textwidth]{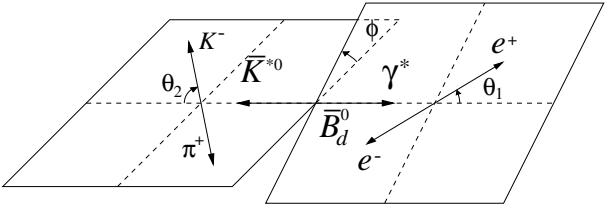}}
\caption{Definition of helicity angles $\theta_1$, $\theta_2$, and
$\phi$, for the decay ${\bar B}_d^0\to {\bar K}^{*0}\,e^+\,e^-$.}
\label{fig1}
\end{figure}

The decay ${\bar B}_d^0\to {\bar K}^{*0}\,e^+\,e^-$, with ${\bar
K}^{*0} \to K^- \pi^+$ on the mass shell~\footnote{This means the
narrow-width approximation for the ${\bar K}^{*0}$ propagator: \
$(k^2 - m_{K^*}^2 + im_{K^*} \Gamma_{K^*})^{-1} \approx -i \pi
\delta(k^2 - m_{K^*}^2) $.}, is completely described by four
independent kinematic variables: the electron-positron pair
invariant-mass squared, $q^2$, and the three angles $\theta_1$,
$\theta_2$, $\phi$. In the helicity frame (Fig.~\ref{fig1}), the
angle $\theta_1\,(\theta_2)$ is defined as the angle between the
directions of motion of $e^+\,(K^-)$ in the $\gamma^*\,({\bar
K}^{*0})$ rest frame and the $\gamma^*\,({\bar K}^{*0})$ in the
${\bar B}_d^0$ rest frame. The azimuthal angle $\phi$ is defined
as the angle between the decay planes of $\gamma^*\to e^+\,e^-$
and ${\bar K}^{*0} \to K^- \pi^+$ in the ${\bar B}_d^0$ rest
frame. The differential decay rate in these coordinates is given
by
\begin{widetext}
\begin{equation}\label{eq:024}
\frac{d^4\,\Gamma}{d\hat{q}^2d\cos\theta_1d\cos\theta_2d\phi}=m_B
\frac{9}{64\,\pi}\sum_{k=1}^{9}a_{k}(q^2)g_{k}(\theta_1,
\theta_2,\phi)\,,
\end{equation}
where the angular terms $g_k$ are defined as
\[g_1=4\sin^2\theta_1\cos^2\theta_2\,,\:
g_2=\left(1+\cos^2\theta_1
-\sin^2\theta_1\cos2\phi\right)\sin^2\theta_2\,,\:
g_3=\left(1+\cos^2\theta_1
+\sin^2\theta_1\cos2\phi\right)\sin^2\theta_2\,,\]
\[g_4=-2\sin^2\theta_1\sin^2\theta_2\sin2\,\phi\,,\:g_5=-\sqrt{2}\sin2\,\theta_1\sin2\,\theta_2\cos\phi\,,
\: g_6=-\sqrt{2}\sin2\,\theta_1\sin2\,\theta_2\sin\phi\,,\]
\[g_7=4\cos\theta_1\sin^2\theta_2\,,\:g_8=-2\sqrt{2}\sin\theta_1\sin2\,\theta_2\cos\phi\,,
\:g_9=-2\sqrt{2}\sin\theta_1\sin2\,\theta_2\sin\phi\,,
\]
and the amplitude terms $a_k$ as
\[a_1=|A_{0}|^2\,, \: a_2=|A_{\|}|^2\,,\: a_3=|A_{\perp }|^2\,,\: a_4={\rm Im}\left(A_{\|}A_{\perp }^*\right)\,,
\: a_5={\rm Re}\left(A_{0}A_{\|}^*\right)\,,\:a_6={\rm
Im}\left(A_{0}A_{\perp }^*\right)\,,\]
\[a_7={\rm Re}\left(A_{\|L}A_{\perp L}^*-A_{\|R}A_{\perp
R}^*\right)\,,\: a_8={\rm Re}\left(A_{0L}A_{\perp
L}^*-A_{0R}A_{\perp R}^*\right)\, ,\:a_9={\rm Im}\left(A_{0L}A_{\|
L}^*-A_{0R}A_{\| R}^*\right)\,, \]
\end{widetext} where $m_B$ is the mass of the $B^0_d$ meson,
$\hat{q}^2\equiv q^2/m_B^2$, and
\begin{equation}
A_i A^*_j \equiv A_{i L}(q^2) A^*_{jL}(q^2)+ A_{iR}(q^2)
A^*_{jR}(q^2) \nonumber \, .
\end{equation}
Here $i,j  = (0, \|, \perp )$, we have neglected the electron mass
$m_e$ and $A_{0L(R)}$, $A_{\|L(R)}$ and $A_{\perp L(R)}$ are the
complex decay amplitudes of the three helicity states in the
transversity basis.

With its rich multidimensional structure, the differential decay
rate in Eq.~(\ref{eq:024}) has sensitivity to various effects
modifying the SM, such as CP violation beyond the
Cabibbo-Kobayashi-Maskawa mechanism and/or right-handed currents.
Given sufficient data, all $a_k$ can, in principle, be completely
measured from the full angular distribution in all three angles
$\theta_1$, $\theta_2$, and $\phi$.

The familiar electron-positron pair invariant-mass spectrum for
${\bar B}_d^0\to {\bar K}^{*0}\,e^+\,e^-$ decay can be recovered
after integration over all angles as
\begin{equation}\label{eq:025}
\frac{d\,\Gamma}{d\hat{q}^2}=m_B\left(|A_0|^2+|A_{\|}|^2+|A_{\perp}|^2
\right)\,.
\end{equation}
The longitudinal and transverse partial widths are given,
respectively, by
\[\frac{d\,\Gamma_0}{d\hat{q}^2}=m_B|A_0|^2\,,\: \quad \quad
\frac{d\,\Gamma_{\rm T}}{d\hat{q}^2}\equiv
\frac{d\,\Gamma_\|}{d\hat{q}^2}+\frac{d\,\Gamma_\perp}{d\hat{q}^2}\,.\]
The fraction of $K^*$ meson polarization is \ $[i=(0, \|, \perp)]$
\[f_i=\frac{d\,\Gamma_i}{d\hat{q}^2}/\frac{d\,\Gamma}{d\hat{q}^2}\,,
\quad \quad \frac{d\,\Gamma_i}{d\hat{q}^2}=m_B|A_i|^2 \,, \]
and $f_{\rm T}=f_\|+f_\perp =1 - f_0$. Integrating
Eq.~(\ref{eq:024}) over the variables $\cos\theta_1$ and $\phi$,
we obtain
\begin{equation}\label{eq:026}
\frac{d^2\,\Gamma}{d\hat{q}^2d\cos\theta_2}=\frac{3}{4}
\frac{d\,\Gamma_{\rm
T}}{d\hat{q}^2}\left(1+\alpha_{K^*}\cos^2\theta_2\right)\,,
\end{equation}
where $\alpha_{K^*}$ is the $K^*$ meson polarization parameter,
$\alpha_{K^*}\equiv 2f_0/f_{\rm T}\,-1\,.$
Integration of Eq.~(\ref{eq:024}) over $\cos\theta_2$ and $\phi$
yields
\[\frac{d^2\,\Gamma}{d\hat{q}^2d\cos\theta_1}=\frac{3}{4}
\frac{d\,\Gamma_0}{d\hat{q}^2}\sin^2\theta_1\]
\begin{equation}\label{eq:027}
+\frac{3}{8} \frac{d\,\Gamma_{\rm
T}}{d\hat{q}^2}(1+\cos^2\theta_1)+\frac{dA_{\rm
FB}}{d\hat{q}^2}\cos\theta_1\,,
\end{equation}
where $A_{\rm FB}$ is forward-backward asymmetry,
\[\frac{d\,A_{\rm FB}}{d\,\hat{q}^2}\equiv\int\limits_{-1}^{1}{\rm sgn}(\cos\theta_1)\frac{d^2\Gamma}
{d\,\hat{q}^2 d\cos\theta_1}\:d\cos\theta_1
\]
\[=\frac{3\,m_B}{2}{\rm
Re}(A_{\parallel\,L}\, A_{\perp\,L}^*-A_{\parallel\,R}\,
A_{\perp\,R}^*)\,,\]
and the normalized forward-backward asymmetries $d{\bar A}_{\rm
FB}/d\hat{q}^2$ and $d\widetilde{A}_{\rm FB}/d\hat{q}^2$ are given
as
\begin{equation} \frac{d{\bar A}_{\rm FB}}{d\hat{q}^2}\equiv\frac{d A_{\rm
FB}}{d\hat{q}^2}/\frac{d \Gamma}{d\hat{q}^2}\,,\qquad
\frac{d\widetilde{A}_{\rm FB}}{d\hat{q}^2}\equiv\frac{d A_{\rm
FB}}{d\hat{q}^2}/\frac{d \Gamma_{\rm T}}{d\hat{q}^2}\,.
\label{eq:0271}
\end{equation}
Finally, the two-dimensional differential decay rate in $q^2$ and
the angle $\phi$ between the lepton and meson planes, after
integration over other variables, takes the form
\begin{equation}\label{eq:028}
\frac{d^2\,\Gamma}{d\hat{q}^2d\phi}=\frac{1}{2\pi}
\frac{d\,\Gamma}{d\hat{q}^2}\left(1+\frac{1}{2}f_{\rm
T}A^{(2)}_{\rm T}\cos2\phi-A_{\rm Im}\sin2\phi\right)\,,
\end{equation}
\begin{equation}\label{eq:0282}
A^{(2)}_{\rm T}\equiv  \frac{ f_\perp-f_\|}{f_{\rm T}}\,, \quad
A_{\rm Im}\equiv m_B{\rm
Im}(A_\|A^*_\perp)/\frac{d\,\Gamma}{d\hat{q}^2}\,,
\end{equation}
\begin{equation}
\widetilde{A}_{\rm Im}\equiv m_B{\rm
Im}(A_\|A^*_\perp)/\frac{d\,\Gamma_{\rm T}}{d\hat{q}^2}\,.
\label{eq:0281}
\end{equation}
%
For $q^2$-integrated quantities we introduce the notation
\[\langle\,X\rangle\equiv \int\limits_{\hat{q}^2_{min}}^{\hat{q}^2_{max}}\frac{dX}{d\hat{q}^2}\:d
\hat{q}^2 \,,\]
where the $X$'s are $\Gamma$ or $\Gamma_i$. Integrated quantities
$\langle f_i\rangle$, $\langle A^{(2)}_{\rm T}\rangle $, and
$\langle A_{\rm Im}\rangle $, which are obtained from the ones
above by integrating the numerator and the denominator separately
over $q^2$, are defined as follows:
\[\langle\,f_i\rangle\equiv
\frac{\langle\,\Gamma_i\rangle}{\langle\,\Gamma\rangle}\,,\: \;
(i=0,\,\perp,\,\parallel)\,, \quad  \langle\,A_{\rm
T}^{(2)}\rangle\equiv
\frac{\langle\,\Gamma_\perp\rangle-\langle\,\Gamma_\parallel\rangle}
{\langle\,\Gamma_\perp\rangle+\langle\,\Gamma_\parallel\rangle}\,,
   \]
\[\frac{d\,\langle\,\Gamma\rangle}{d\,\phi}=
\frac{\langle\,\Gamma\rangle}{2\pi}\left(1+\frac{1}{2}\langle
f_{\rm T}\rangle\langle A_{\rm T}^{(2)}\rangle\cos2\phi-\langle
A_{\rm Im}\rangle\sin2\phi\right)\,,\]
\[
\langle\,A_{\rm Im}\rangle\equiv m_B\frac{\langle {\rm
Im}A_{\parallel}\,
A_{\perp}^*\rangle}{\langle\,\Gamma\rangle}\,,\] \[ \langle {\rm
Im}A_{\parallel}\,
A_{\perp}^*\rangle\equiv\int\limits_{\hat{q}^2_{min}}^{\hat{q}^2_{max}}
{\rm Im}(A_{\parallel}\, A_{\perp}^*)d\hat{q}^2\,.
\]

\subsection{ \label{subsec:transversity}
Transversity amplitudes}

The nonresonant amplitudes follow from the matrix element of the
${\bar B}_d^0 (p)\to {\bar
K}^{*0}(k,\epsilon)\,e^+(q_+)\,e^-(q_-)$ process in
Eq.~(\ref{eq:005}),
\begin{widetext}
\[A_{0L,R}^{\rm NR}=-\frac{N
\hat{\lambda}^{1/4}}{2\hat{m}_{K^*}}\Biggl((C_{9V}^{\rm eff}\mp
C_{10A})\left((1-\hat{q}^2-\hat{m}_{K^*}^2)(1+\hat{m}_{K^*})\,
A_1(q^2)-\hat{\lambda} \frac{A_2(q^2)}{1+\hat{m}_{K^*}}\right)\]
\begin{equation}\label{eq:029}
+2(\hat{m}_b-\hat{m}_s)\,C_{7\gamma}^{\rm
eff}\left((1-\hat{q}^2+3\hat{m}_{K^*}^2)\,T_2(q^2)-
\frac{\hat{\lambda}}{1-\hat{m}_{K^*}^2}T_3(q^2)\right)\Biggr)\,,
\end{equation}
\begin{equation}\label{eq:030}
A_{\|L,R}^{\rm NR}=N(1-\hat{m}_{K^*}^2)\sqrt{2\hat{q}^2}\,
\hat{\lambda}^{1/4}\Bigl((C_{9V}^{\rm eff}\mp
C_{10A})\frac{A_1(q^2)}{1-\hat{m}_{K^*}}+
2\frac{\hat{m}_b-\hat{m}_s}{\hat{q}^2}\,C_{7\gamma}^{\rm
eff}\,T_2(q^2)\Bigr)\,,
\end{equation}
\begin{equation}\label{eq:031}
A_{\perp L,R}^{\rm NR}=-N\sqrt{2\hat{q}^2}\,
\hat{\lambda}^{3/4}\Bigl((C_{9V}^{\rm eff}\mp
C_{10A})\frac{V(q^2)}{1+\hat{m}_{K^*}}+
2\frac{\hat{m}_b+\hat{m}_s}{\hat{q}^2}\,C_{7\gamma}^{\rm
eff}\,T_1(q^2)\Bigr)\,.
\end{equation}
\end{widetext}
In the above formulas the definition  \ $\hat{m}_{K^*}\equiv
m_{K^*}/m_B$, $\hat{\lambda}\equiv\lambda(1,\hat{q}^2,
\hat{m}_{K^*}^2)=(1-\hat{q}^2)^2-2(1+\hat{q}^2)\hat{m}_{K^*}^2+\hat{m}_{K^*}^4$,
$\hat{m}_b\equiv \overline{m}_b(\mu)/m_B$, $\hat{m}_s\equiv
\overline{m}_s(\mu)/m_B$ are used, where $m_{K^*}$ is the mass of
the $K^{*0}$ meson, and
\[N=|V_{tb}V_{ts}^*|\frac{G_F m_B^2 \alpha_{\rm em}}{32 \,\pi^2 \sqrt{3\,
\pi}}\,.\]
The transversity amplitudes in Eqs.~(\ref{eq:029})--(\ref{eq:031})
take a particularly simple form in the heavy-quark and
large-energy limit. In fact, exploiting the form factor relations
in Eqs.~(\ref{eq:017})--(\ref{eq:023}), we obtain
\begin{widetext}
\[A_{0L,R}^{\rm NR}=-N \hat{\lambda}^{1/4}\Biggl((C_{9V}^{\rm
eff}\mp
C_{10A})\left(\hat{m}_{K^*}(1+\hat{q}^2-\hat{m}_{K^*}^2)\,\xi_\perp(q^2)
+\frac{\hat{\lambda}}{2\hat{m}_{K^*}}\,\xi_\|(q^2) \right)\]
\begin{equation}\label{eq:0291}
+2(\hat{m}_b-\hat{m}_s)\,C_{7\gamma}^{\rm
eff}\left(2\,\hat{m}_{K^*}\,\xi_\perp(q^2)+
\frac{\hat{\lambda}}{2\hat{m}_{K^*}}\,\xi_\|(q^2)\right)\Biggr)\,,
\end{equation}
\begin{equation}\label{eq:0301}
A_{\|L,R}^{\rm NR}=N\sqrt{2\hat{q}^2}\,
\hat{\lambda}^{1/4}\Bigl(\left(C_{9V}^{\rm eff}\mp
C_{10A}\right)\left(1-\hat{q}^2+\hat{m}_{K^*}^2\right)+
2\frac{\hat{m}_b-\hat{m}_s}{\hat{q}^2}\left(1-\hat{q}^2-\hat{m}_{K^*}^2\right)
\,C_{7\gamma}^{\rm eff}\Bigr)\,\xi_\perp(q^2)\,,
\end{equation}
\begin{equation}\label{eq:0311}
A_{\perp L,R}^{\rm NR}=-N\sqrt{2\hat{q}^2}\,
\hat{\lambda}^{3/4}\Bigl(C_{9V}^{\rm eff}\mp C_{10A} +
2\frac{\hat{m}_b+\hat{m}_s}{\hat{q}^2}\,C_{7\gamma}^{\rm
eff}\Bigr)\,\xi_\perp(q^2)\,.
\end{equation}
From inspection of these formulas we infer the following features.
The amplitudes (\ref{eq:0301}) and (\ref{eq:0311}) are expressed
through the one form factor $\xi_\perp(q^2)$. The observables
$A^{(2)}_{\rm T}$, \ $\widetilde{A}_{\rm Im}$, and
$d\widetilde{A}_{\rm FB}/d\hat{q}^2$ do not depend on the
functional form of the form factor $\xi_\perp(q^2)$, and therefore
they can be used to study the Wilson coefficients.

In the region ${q}^2\lesssim {m}_{K^*}^2=0.803$ GeV$^2$, the
transversity amplitudes (\ref{eq:0291})--(\ref{eq:0311}) take the
form
\[A_{0L,R}^{\rm NR}\approx -\frac{N}{2\hat{m}_{K^*}} \Biggl((C_{9V}^{\rm
eff}\mp C_{10A})\left(2\hat{m}_{K^*}^2\,\xi_\perp(q^2)
+\left(1-\frac{5}{2}\left(\hat{q}^2+\hat{m}_{K^*}^2\right)\right)
\xi_\|(q^2)\right)\]
\begin{equation}\label{eq:0312}
+2(\hat{m}_b-\hat{m}_s)\,C_{7\gamma}^{\rm
eff}\left(4\hat{m}_{K^*}^2\,\xi_\perp(q^2)
+\left(1-\frac{5}{2}\left(\hat{q}^2+\hat{m}_{K^*}^2\right)\right)
\xi_\|(q^2)\right)\Biggr)\,,
\end{equation}
\begin{equation}\label{eq:0313}
A_{\|L,R}^{\rm NR}\approx N\sqrt{\frac{2}{\hat{q}^2}}
\Biggl(\left(C_{9V}^{\rm eff}\mp C_{10A}\right)\hat{q}^2+
2\left(\hat{m}_b-\hat{m}_s\right)\left(1-\frac{3}{2}\left(\hat{q}^2+
\hat{m}_{K^*}^2\right)\right)\,C_{7\gamma}^{\rm
eff}\Biggr)\,\xi_\perp(q^2)\,,
\end{equation}
\begin{equation}\label{eq:0314}
A_{\perp L,R}^{\rm NR}\approx -N\sqrt{\frac{2}{\hat{q}^2}}
\Biggl(\left(C_{9V}^{\rm eff}\mp C_{10A}\right)\hat{q}^2+
2\left(\hat{m}_b+\hat{m}_s\right)\left(1-\frac{3}{2}\left(\hat{q}^2+
\hat{m}_{K^*}^2\right)\right)\,C_{7\gamma}^{\rm
eff}\Biggr)\,\xi_\perp(q^2)\,.
\end{equation}
\end{widetext}

It follows from these equations that, in the region of very small
invariant masses, namely, ${q}^2 \ll {m}_{K^*}^2$, the asymmetry
$A^{(2)}_{\rm T}$ in Eq.~(\ref{eq:0282}) takes the simple form
\begin{equation}\label{eq:0315}
A^{(2)}_{\rm T}\approx \frac{2 {m}_s}{{m}_b}\,.
\end{equation}
This result is in agreement with the well-known fact that, in the
SM for $m_s=0$ in a naive factorization, $A^{(2)}_{\rm
T}=0$~\cite{Kruger:2005}.

In some extensions of the SM, such as the left-right model and the
unconstrained supersymmetric SM, there are right-handed currents
in the matrix element, with the magnitude determined by the
coupling $C_{7\gamma}^{\prime\,{\rm eff}}$ (see, e.g.,
Ref.~\cite{Kruger:2005}). In this case the asymmetry $A^{(2)}_{\rm
T}$ is written as
\begin{equation}\label{eq:0316}
A^{(2)}_{\rm T}\approx \frac{2C_{7\gamma}^{\prime\,\rm
eff}\,C_{7\gamma}^{\rm eff}}{(C_{7\gamma}^{\rm
eff})^2+(C_{7\gamma}^{\prime\,{\rm eff}})^2}.
\end{equation}

\subsection{\label{subsec:resonances}
 Resonant contribution}

\begin{figure}[tbh]
\centerline{\includegraphics[width=.45\textwidth]{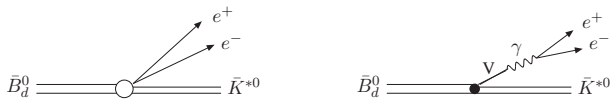}}
\caption{Nonresonant and resonant contributions to the decay
amplitude. } \label{fig:resonances}
\end{figure}

Next, we implement the effects of LD contributions from the decays
${\bar B}_d^0\to {\bar K}^{*0}\,V$, where $V=\rho^0\,,\omega\,,
\phi$ mesons, followed $V\to e^+\,e^-$ in the decay ${\bar
B}_d^0\to {\bar K}^{*0}\,e^+\,e^-$~(see
Fig.~\ref{fig:resonances}). Using the vector-meson dominance
concept we obtain the amplitude including nonresonant and resonant
parts,
\[A_{\lambda L,R}=A_{\lambda L,R}^{\rm NR}\]
\begin{equation}\label{eq:032}
+\sum_{V}\frac{c_V e^{i\delta_V}}{D_V(\hat{q}^2)}
\Bigl(\frac{\lambda(1,\hat{q}^2,
\hat{m}_{K^*}^2)}{\lambda(1,\hat{m}_V^2,
\hat{m}_{K^*}^2)}\Bigr)^{1/4}\frac{\hat{m}_V}{\sqrt{\hat{q}^2}}\,
h_\lambda^V\,,
\end{equation}
\begin{widetext}
\begin{equation} c_V= {\rm sgn}(Q_V)\Bigl({\rm Br}(V\to
e^+\,e^-){\rm Br}({\bar B}_d^0\to {\bar K}^{*0}\,V){\hat m}_V{\hat
\Gamma}_V/(2\pi m_B\tau_B)\Bigr)^{1/2} \, , \label{eq:0321}
\end{equation}
\end{widetext}
where $\lambda= (0, \|, \perp)$ and
\[ D_V(\hat{q}^2) = \hat{q}^2 - \hat{m}_V^2 +
i\hat{m}_V \hat{\Gamma}_V (\hat{q}^2) \] is the usual Breit-Wigner
function for the $V$ meson resonance shape with the
energy-dependent width ${\Gamma}_V ({q}^2)$ \ [$\hat{\Gamma}_V
(\hat{q}^2) = {\Gamma}_V ({q}^2) /m_B $]. In Eq.~(\ref{eq:0321}) \
$Q_V$ is the effective electric charge of the quarks in the vector
meson $V$ ($Q_\rho=1/\sqrt{2}$, $Q_\omega=1/\sqrt{18}$,
$Q_\phi=-1/3$), $\hat{m}_V\equiv m_V/m_B$, $\hat{\Gamma}_V\equiv
\Gamma_V/m_B$, $m_V(\Gamma_V)$ is the mass (width) of a $V$ meson,
${\rm Br}(\ldots)$ is the branching ratio, the $\tau_B$ is the
lifetime of a $B$ meson. In addition, $h_\lambda^V$
($\lambda=0\,,\|\,,\perp$) are the complex amplitudes for ${\bar
B}_d^0\to {\bar K}^{*0}\,V$ decay processes of the three helicity
states in the transversity basis with the normalization condition
$|h_0^V|^2+|h_\|^V|^2+|h_\perp^V|^2=1$, and $\delta_V$ is the
phase of the resonant amplitude relative to the phase of the
nonresonant one.

Parameters of the vector resonances are presented in
Table~\ref{tab:param1}. The energy-dependent width for the $\rho$
meson is chosen as \cite{PDG:2008}
 \begin{equation}
\Gamma_{\rho}(q^2) = \Gamma_\rho \frac{m_\rho }{\sqrt{q^2}} \frac{
k^3 }{k_0^3 }  \, \frac{1+ r^2 k_0^2}{1+ r^2 k^2} \, \Theta
\bigl(q^2 - 4 m_\pi^2 \bigr)
 \label{eq:0322}
 \end{equation}
where $k = (q^2/4 - m_\pi^2)^{1/2}$, \ $k_0 = (m_\rho^2/4 -
m_\pi^2)^{1/2}$ and the parameter $r= 2.5$ GeV$^{-1}$
\cite{Achasov:2003}, $\Theta (x) =1$ for $x \ge 0$, and $\Theta
(x) =0$ otherwise.

For the $\omega$ meson we take the energy dependence of the width
in the form
\begin{table}[th]
\caption{Mass, width, the leptonic branching ratio of the
$\rho^0$, $\omega$, and $\phi$ mesons \cite{PDG:2008} .}
\label{tab:param1}
\begin{center}
\begin{tabular}{c c c c}
\hline \hline $V$ & $m_V\,{\rm (GeV)}$ &$\Gamma_V\,{\rm (GeV)}$&${\rm Br}(V\to e^+\,e^-)$  \\
\hline
$\rho^0$   & $0.77549$ & $0.1462$ & $4.71\times 10^{-5}$  \\
$\omega$   &$0.78265$  & $0.00849$ & $7.16\times 10^{-5}$   \\
$\phi$     &$1.019455$  & $0.00426$ & $2.97\times 10^{-4}$  \\
\hline \hline
\end{tabular}
\end{center}
\end{table}

\begin{widetext}
\[\Gamma_{\omega}(q^2) = \Gamma_\omega \Bigl[ {\rm Br}(\omega \to
3\pi) \, \Theta \bigl(q^2 - 9 m_\pi^2 \bigr) + {\rm Br}(\omega \to
\pi^0 \gamma) \, \Theta \bigl(q^2 - m_\pi^2 \bigr) + {\rm
Br}(\omega \to 2\pi) \, \Theta \bigl(q^2 - 4 m_\pi^2 \bigr)
\Bigr],\] where the branching ratios are \ ${\rm Br}(\omega \to
3\pi)=89.2$\%, \ ${\rm Br}(\omega \to \pi^0 \gamma)=8.92$\%, and
${\rm Br}(\omega \to 2\pi)=1.53$\% \cite{PDG:2008}, and for the
$\phi$ meson,
\[\Gamma_{\phi}(q^2) = \Gamma_\phi \Bigl[ {\rm Br}(\phi \to K^+
K^-) \, \Theta \bigl(q^2 - 4 m_{K^{\pm}}^2 \bigr) + {\rm Br}(\phi
\to K^0 \bar{K}^0 ) \, \Theta \bigl(q^2 - 4 m_{K^0}^2 \bigr)\]\[ +
{\rm Br}(\phi \to 3 \pi) \, \Theta \bigl(q^2 - 9 m_\pi^2 \bigr) +
{\rm Br}(\phi \to \eta \gamma ) \, \Theta \bigl(q^2 - m_\eta^2
\bigr) \Bigr],\] with the branching ratios ${\rm Br}(\phi \to K^+
K^-)=49.2 $\%, \ ${\rm Br}(\phi \to K^0 \bar{K}^0 ) =34.0$\%, \
${\rm Br}(\phi \to 3 \pi)=15.25$\%, and ${\rm Br}(\phi \to \eta
\gamma )=1.304$\% \cite{PDG:2008}. \end{widetext}

\begin{figure*}
\centerline{\includegraphics[width=.9\textwidth]{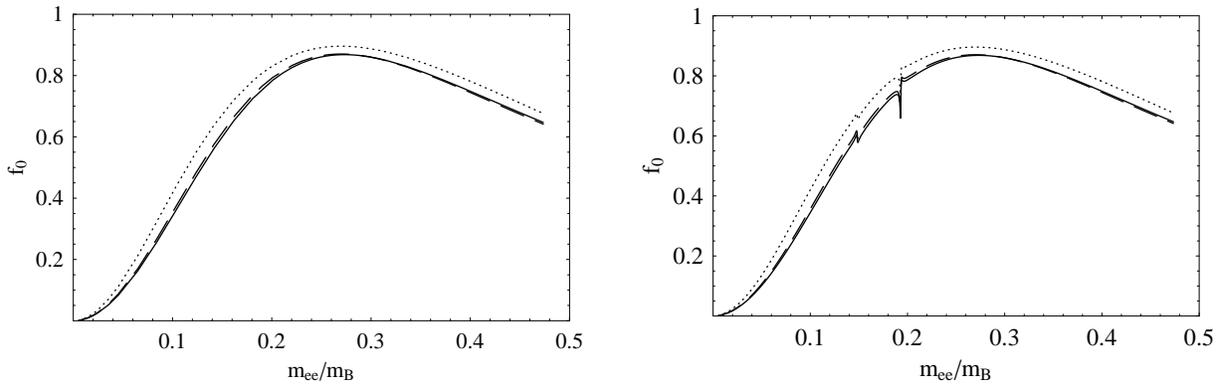}}
\caption{Longitudinal polarization fraction of the $K^*$ meson as
a function of $m_{ee}/m_B$. Left (right) panel corresponds to the
calculation without (with) resonances taken into account. The mass
of the strange quark is $m_s=79$ MeV. The dashed lines correspond
to the form factor model from~\cite{Ball:2005}, the dotted lines
correspond to the model from~\cite{Ali:2000}, and the solid lines
are calculated according to Eqs.~(\ref{eq:014})--(\ref{eq:016})
and (\ref{eq:0231}). } \label{fig:f0}
\end{figure*}

In order to calculate the resonant contribution to the amplitude
of the ${\bar B}_d^0\to {\bar K}^{*0}\,e^+\,e^-$ decay, one has to
know the amplitudes of the decays ${\bar B}_d^0\to {\bar K}^{*0}\,
\rho$, \ ${\bar B}_d^0\to {\bar K}^{*0}\,\omega$, and ${\bar
B}_d^0\to {\bar K}^{*0}\,\phi$. Unfortunately, at present only the
amplitude of the ${\bar B}_d^0\to {\bar K}^{*0}\,\phi$ decay is
known from experiment~\cite{hfag:2009}; therefore, in our estimate
we use the amplitudes of the ${\bar B}_d^0\to {\bar K}^{*0}\,
\rho$ and ${\bar B}_d^0\to {\bar K}^{*0}\,\omega$ decays from the
theoretical prediction~\cite{Chen:2006}. The absolute values and
phase of the normalized decay amplitudes $h_\lambda^V$ are shown
in Table~\ref{tab:ampl}.

\begin{table}
\caption{Branching ratios \cite{hfag:2009} and decay amplitudes
for ${\bar B}_d^0\to {\bar K}^{*0}\,\rho^0$ \cite{Chen:2006},
${\bar B}_d^0\to {\bar K}^{*0}\,\omega$ \cite{Chen:2006}, and
${\bar B}_d^0\to {\bar K}^{*0}\,\phi$ \cite{hfag:2009}.}
\label{tab:ampl}
\begin{center}
\begin{tabular}{c c c c}
\hline \hline Mode & ${\bar K}^{*0}\,\rho^0$ &${\bar K}^{*0}\,\omega$&${\bar K}^{*0}\,\phi$ \\
\hline ${\rm Br}({\bar B}_d^0\to {\bar K}^{*0}\,V)$&$3.4\times
10^{-6}$&$2.0\times 10^{-6}$&$9.8\times 10^{-6}$\\
$|h_0^V|^2$       & $0.70$ & $0.75$ & $0.480$ \\
$|h_\perp^V|^2$   &$0.14$  & $0.12$ & $0.241$\\
${\rm arg}(h_\|^V/h_0^V)$ (rad)&$1.17$  & $1.79$ & $2.40$\\
${\rm arg}(h_\perp^V/h_0^V)$ (rad)&$1.17$& $1.82$ & $2.39$ \\
\hline \hline
\end{tabular}
\end{center}
\end{table}

\begin{figure*}
\centerline{\includegraphics[width=.9\textwidth]{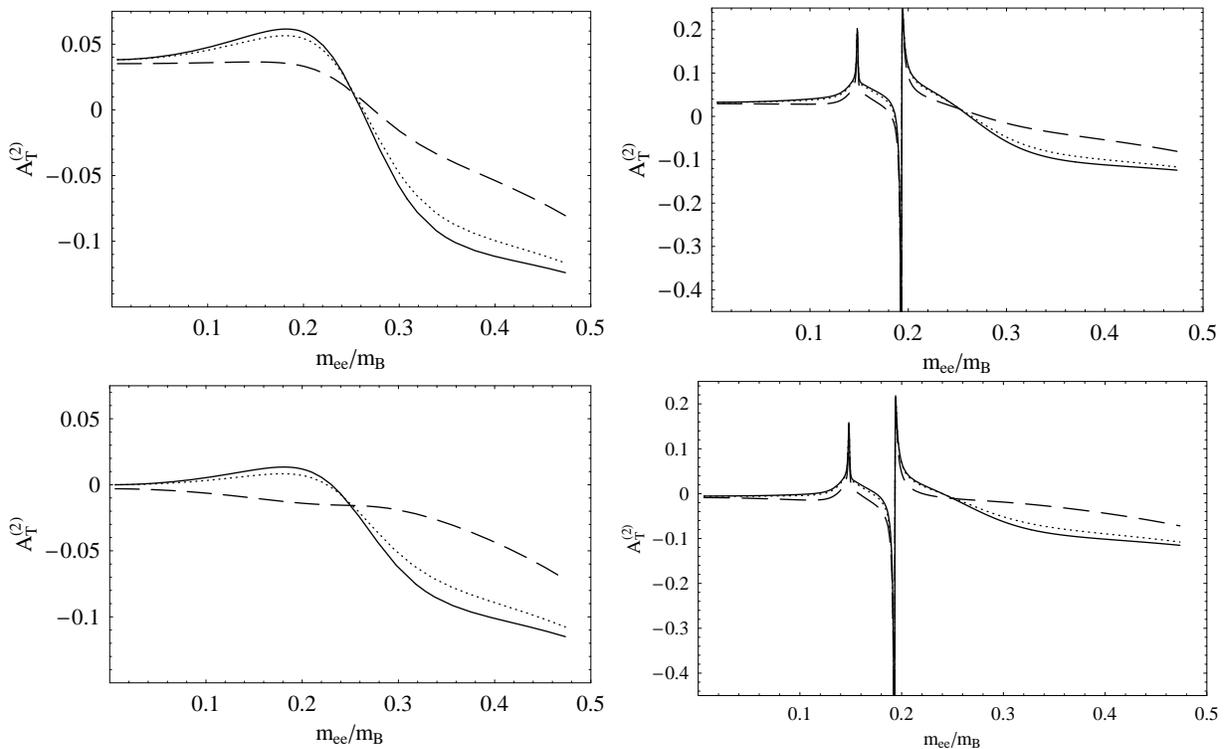}}
\caption{Transverse asymmetry as a function of $m_{ee}/m_B$. Left
(right) panels correspond to the calculation without (with)
resonances taken into account. Top (bottom) panels correspond to
the calculation with mass $m_s=79$ MeV  ($m_s=0$). The lines are
defined as in Fig.~\ref{fig:f0}.} \label{fig:AT2}
\end{figure*}

\begin{figure*}
\centerline{\includegraphics[width=.9\textwidth]{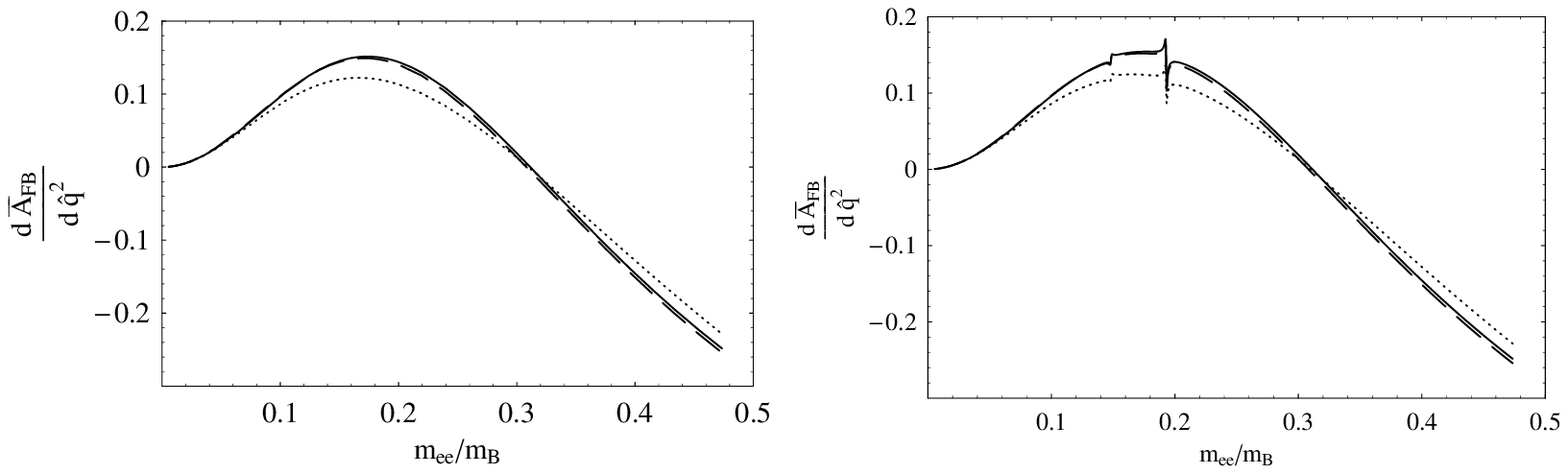}}
\caption{The normalized forward-backward asymmetry $d{\bar A}_{\rm
FB}/d\hat{q}^2$ as a function of $m_{ee}/m_B$. Left (right) panel
corresponds to the calculation without (with) resonances taken
into account. The mass of the strange quark is $m_s=79$ MeV. The
lines are defined as in Fig.~\ref{fig:f0}. } \label{fig:FB}
\end{figure*}

\begin{figure*}
\centerline{\includegraphics[width=.9\textwidth]{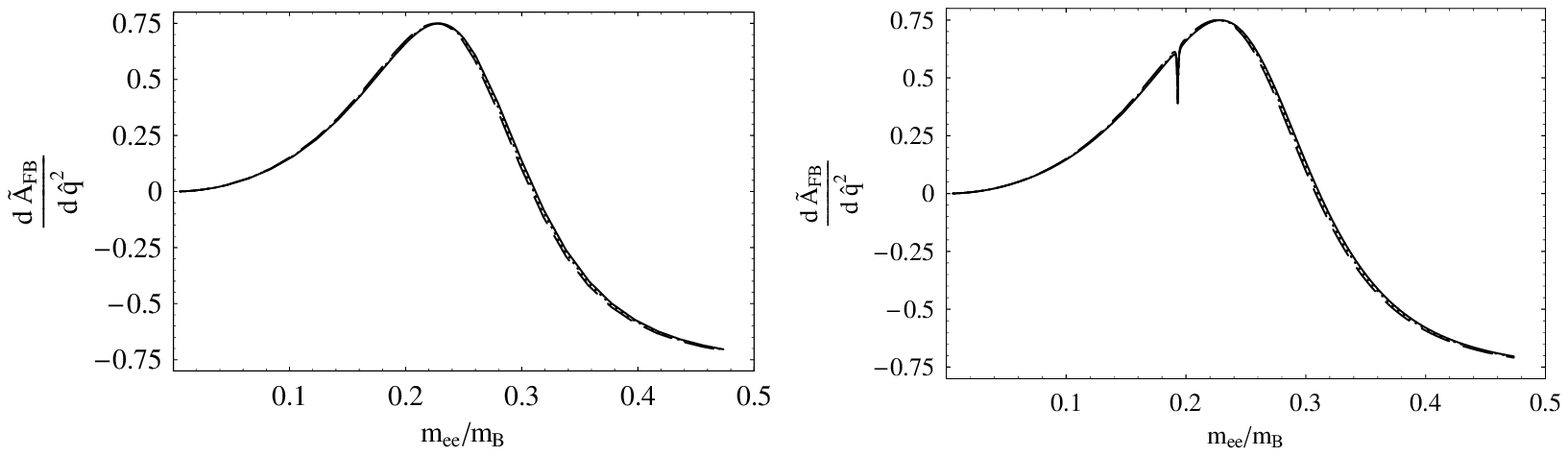}}
\caption{The normalized forward-backward asymmetry
$d\widetilde{A}_{\rm FB}/d\hat{q}^2$ as a function of
$m_{ee}/m_B$. Left (right) panel corresponds to the calculation
without (with) resonances taken into account. The mass of the
strange quark is $m_s=79$ MeV. The lines are defined as in
Fig.~\ref{fig:f0}. } \label{fig:FBmod}
\end{figure*}

\section{ \label{sec:results}
 Results of the calculation for the ${\bar B}_d^0\to {\bar
K}^{*0}\,e^+\,e^-$ decay and a discussion}

The parameters of the model are indicated in
Table~\ref{tab:param2}. The SM Wilson coefficients at the scale
$\mu=4.8$ GeV to NNLO accuracy \cite{Altmannshofer:2009} are given
in Table~\ref{tab:Wilson}.

\begin{table}[t]
\caption{The numerical input used in our analysis.}
\label{tab:param2}
\begin{center}
\begin{tabular}{ll}
\hline \hline
$|V_{tb}V_{ts}^*|=0.0407$ & $G_F=1.16637\times 10^{-5}\, {\rm GeV^{-2}}$\\

$\mu=m_b=4.8\, {\rm GeV}$   & $\alpha_{\rm em}=1/137$\\

$m_c=1.4\, {\rm GeV}$       &$m_B=5.27953\, {\rm GeV}$ \\

$\overline{m}_b(\mu)=4.14\, {\rm GeV}$     &$\tau_B=1.530\, {\rm ps}$ \\

$\overline{m}_s(\mu)=0.079\, {\rm GeV}$     &$m_{K^*}=0.896\, {\rm GeV}$\\

\hline \hline
\end{tabular}
\end{center}
\end{table}

\begin{table}[th]
\caption{The SM Wilson coefficients at the scale $\mu=4.8$\,GeV,
to NNLO accuracy. Input: $\alpha_s(m_W)=0.120$,
$\alpha_s(\mu)=0.214$, obtained from $\alpha_s(m_Z)=0.1176$
\cite{PDG:2008}, using three-loop evolution,
${\overline{m}}_t({\overline{m}}_t)=162.3\,$GeV, $m_W = 80.4\,$GeV
and $\sin^2\theta_W = 0.23$. } \label{tab:Wilson}
\begin{center}
\begin{tabular}{ccccc}
\hline \hline
$\bar C_1(\mu)$ & ${\bar C}_2(\mu)$ &${\bar C}_3(\mu)$ &${\bar C}_4(\mu)$ &${\bar C}_5(\mu)$ \\
$-0.128$ & $1.052$ & $0.011$ & $-0.032$ &$0.009$  \\
${\bar C}_6(\mu)$ & $C_{7\gamma}^{\rm eff}(\mu)$ & $C_{8 g}^{\rm eff}(\mu)$ &$C_{9{\rm V}}(\mu)$&$C_{10{\rm A}}(\mu)$\\
$-0.037$ & $-0.304$ &$-0.167$& $4.211$&$-4.103$  \\
\hline \hline
\end{tabular}
\end{center}

\end{table}

\subsection{\label{subsec:differential}
Invariant-mass distributions}

In Figs.~\ref{fig:f0}-\ref{fig:FBmod} we present results for the
invariant-mass dependence of various observables for the ${\bar
B}_d^0\to {\bar K}^{*0}\,e^+\,e^-$ decay. The upper limit of the
invariant-mass region, 2.5 GeV, is taken to exclude the
contribution from $J/\psi (1S)$ and higher resonances. Of course,
the presented results may depend on the relative phase $\delta_V$
in Eq.~(\ref{eq:032}). In the figures below we choose this phase
to be equal for all resonances $\rho,\, \omega$, and $\phi$, and
equal to zero.

The polarization parameter $f_0$ for $K^*$ is shown in
Fig.~\ref{fig:f0} (recall that the transverse polarization
fraction is related to $f_0$ through $f_T=1-f_0$). The resonances
$\rho,\, \omega$, and $\phi$ show up as small irregularities on
the smooth curves. These parameters are calculated for the mass of
the strange quark $m_s=79$ MeV. The results of the calculation for
$m_s=0$ are not shown because they are indistinguishable from the
curves for $m_s=79$ MeV. One can also see a certain dependence on
the choice of form factors.

The asymmetry $A_{\rm T}^{(2)}$ in the SM without and with
resonant contributions is presented in Fig.~\ref{fig:AT2}. This
observable turns out to be sensitive to all ingredients of the
model. The dependence on the form factor model is quite essential,
especially if resonances are not included (left-hand side). The
addition of the resonances drastically changes this observable
(see the right-hand side of Fig.~\ref{fig:AT2}). In particular,
the $\phi$ meson contribution is very pronounced.

$ A_{\rm T}^{(2)}$ is also sensitive to the mass of the strange
quark (compare the top and bottom panels of Fig.~\ref{fig:AT2}).
In particular, comparing the top and bottom panels (with
resonances), one concludes that, at an invariant mass below 0.5
GeV (where $q^2 \ll m_{K^*}^2 $), for $m_s=0$ $A_T^{(2)}$
vanishes, while for $m_s=79$ MeV $ A_{\rm T}^{(2)}$ is about 0.04.
This value is in agreement with Eq.~(\ref{eq:0315}). Note that
this asymmetry is, in general, an important observable to study
effects of NP~\cite{Kruger:2005}. Indeed, comparison of $ A_{\rm
T}^{(2)}$ for $m_s=0$ and $m_s \ne 0$ demonstrates the effect of
the ``wrong'' helicity transition $b_L \to s_R + \gamma_R$. In the
SM this effect appears to be small, being proportional to the
ratio $2 m_s/m_b$, while in some extensions of the SM it can reach
bigger values depending on the coefficient
$C_{7\gamma}^{\prime\,{\rm eff}}$  in Eq.~(\ref{eq:0316}) \ (see,
e.g., the estimates in~\cite{Kruger:2005}). It also follows that
the effect of the nonzero mass $m_s$ is bigger than the
uncertainty related to the chosen model for the transition form
factors.

In general, theoretical uncertainties of the nonresonant
amplitudes arise due to the choice of the renormalization scale
$\mu$ (the scale at which the Wilson coefficients, $\alpha_s$, and
$\overline{\rm{ MS}}$ masses are calculated), the ratio $m_c/m_b$,
and some other uncertainties \cite{BFS:2001}. There are also
corrections of the order $\Lambda_{QCD}/m_b $ which are evaluated
in Refs.~\cite{Altmannshofer:2009,Egede:2010}. While it is assumed
in \cite{Altmannshofer:2009} that the main part of the
$\Lambda_{QCD}/m_b $ corrections is inside the QCD form factors,
the authors of \cite{Egede:2010} explicitly include these
corrections in the amplitudes $A_{\lambda L,R}^{\rm NR}$. For an
estimate of the theoretical error of the calculation of the
asymmetry $ A_{\rm T}^{(2)}$, we can use the result of
\cite{Egede:2010}, in which the $\Lambda_{QCD}/m_b $ corrections
to each spin amplitude are estimated to be $\pm 10\%$.  That leads
to the total uncertainty $ A_{\rm T}^{(2)}$ about $\pm 0.05$, in
the SM with $m_s=0$ (see Fig.~14 in \cite{Egede:2010}). In view of
this, the effect of the nonzero mass of the strange quark observed
in Fig.~\ref{fig:AT2} may be overshadowed by theoretical
uncertainties, although this aspect requires further
investigation.

Finally, in Figs.~\ref{fig:FB} and \ref{fig:FBmod} we show the
normalized forward-backward asymmetries in Eq.~(\ref{eq:0271}).
Usually, for the normalized forward-backward asymmetry the
quantity $d{\bar A}_{\rm FB}/d\hat{q}^2$ is chosen. Along with
this one can define the forward-backward asymmetry
$d\widetilde{A}_{\rm FB}/d\hat{q}^2$, normalized in a different
way [cf. Eq.~(\ref{eq:0271})]. Comparing both figures we see that
the latter asymmetry in Fig.~\ref{fig:FBmod} has interesting
properties: i) it is almost independent of the form factor model,
and ii) it may reach values up to $\pm$0.75 which are much larger
than the maximal values taken by the asymmetry in
Fig.~\ref{fig:FB}. These properties, in our opinion, make
$d\widetilde{A}_{\rm FB}/d\hat{q}^2$ a convenient observable for
experimental study. Note that both these asymmetries change
insignificantly when going from $m_s=0$ to $m_s=79$ MeV.


\subsection{\label{subsec:integrated}
Observables integrated over $q^2$ }

In Tables~\ref{tab:sm1} -- \ref{tab:smLD2} we present results of
the calculation of various observables in the framework of the SM,
integrated over $q^2$. Two regions of $e^+ e^-$ invariant mass
$m_{ee} \equiv \sqrt{q^2}$ are considered: (a) 0.030 GeV $<
m_{ee}<$ 1 GeV and (b) 0.5 GeV $< m_{ee}<$ 1 GeV. These intervals
are selected because they turn out to be convenient for future
experiments being planned at the LHCb (see
Ref.~\cite{Lefrancois:2009}). In particular, the limit 0.030 GeV
for interval (a) is taken because at lower masses, $m_{ee} <$
0.030 GeV, it is difficult to define the plane of the lepton pair.
When selecting interval (b) we took into account that the
resolution on the $\phi$ angle in Fig.~\ref{fig1}, according to
the analysis of \cite{Lefrancois:2009}, for $m_{ee}>$ 0.5 GeV is
considerably better than the resolution for $m_{ee}< $ 0.5 GeV. In
addition, in region (b) the vector resonances $\rho, \omega, \phi$
are expected to show up most prominently.

\begin{table}[t]
\caption{Predictions of the SM for the integrated branching ratio
$\tau_B\,\langle\Gamma\rangle$, the polarization parameters
$\langle\,f_i\rangle$, and the asymmetries $\langle\,A_{\rm
T}^{(2)}\rangle$, $\langle\,A_{\rm Im}\rangle$ with the
integration boundaries $0.030\,{\rm GeV}\leq m_{ee} \leq 1\,{\rm
GeV} $. The contribution of the resonances $\rho, \omega, \varphi$
is not included. FF stands for the form factor model chosen
according to \cite{Ali:2000}, \cite{Ball:2005} and
Eqs.~(\ref{eq:014})--(\ref{eq:016}) and (\ref{eq:0231}). }
\label{tab:sm1}
\begin{center}
\begin{tabular}{|c|c|c|c|c|c|c|}
\hline  &
\multicolumn{3}{|c|}{$m_s=79\,{\rm MeV}$}& \multicolumn{3}{|c|}{$m_s=0$}\\
\cline{2-7}  &FF\cite{Ali:2000} & FF\cite{Ball:2005} &
 FF  & FF\cite{Ali:2000}& FF\cite{Ball:2005}
 & FF \\
\hline
$\tau_B\,\langle\Gamma\rangle\times10^{7}$ &$1.92$&$1.79$&$1.99$&$1.92$&$1.79$&$1.99$\\
$\langle\,f_0\rangle$ &$0.25$&$0.20$&$0.19$&$0.25$&$0.20$&$0.19$\\
$\langle\,f_\perp\rangle$ &$0.39$&$0.41$&$0.42$&$0.38$&$0.40$&$0.41$\\
$\langle\,f_\parallel\rangle$ &$0.36$&$0.39$&$0.39$&$0.38$&$0.40$&$0.40$ \\
$\langle\,A_{\rm T}^{(2)}\rangle\times10^{2}$&$4.1$&$3.5$&$4.2$&$0.1$&$-0.4$&$0.2$ \\
$\langle\,A_{\rm Im}\rangle\times10^{5}$&$2.$&$1.$&$3.$&$1.$&$0.$&$2.$ \\
\hline
\end{tabular}
\end{center}
\end{table}

\begin{table}[t]
\caption{Same as Table \ref{tab:sm1} but with the integration
boundaries $0.5\,{\rm GeV}\leq m_{ee}\leq 1\,{\rm GeV} $.}
\label{tab:sm2}
\begin{center}
\begin{tabular}{|c|c|c|c|c|c|c|}
\hline &
\multicolumn{3}{|c|}{$m_s=79\,{\rm MeV}$}& \multicolumn{3}{|c|}{$m_s=0$}\\
\cline{2-7}  &FF\cite{Ali:2000}& FF\cite{Ball:2005} &
 FF & FF\cite{Ali:2000} & FF\cite{Ball:2005}& FF \\
\hline
$\tau_B\,\langle\Gamma\rangle\times10^{8}$ &$5.7$&$4.8$&$5.2$&$5.7$&$4.8$&$5.2$\\
$\langle\,f_0\rangle$ &$0.62$ & $0.56$&$0.55$&$0.62$&$0.56$&$0.55$\\
$\langle\,f_\perp\rangle$ &$0.20$&$0.23$&$0.24$&$0.19$&$0.22$&$0.23$\\
$\langle\,f_\parallel\rangle$ &$0.18$&$0.21$&$0.22$&$0.19$&$0.22$&$0.23$ \\
$\langle\,A_{\rm T}^{(2)}\rangle\times10^{2}$&$5.0$&$3.6$&$5.4$&$0.6$&$-0.9$&$0.9$ \\
$\langle\,A_{\rm Im}\rangle\times10^{5}$&$5.$&$2.$&$8.$&$3.$&$-1.$&$5.$ \\
\hline
\end{tabular}
\end{center}
 \end{table}

\begin{table}[t]
\caption{Predictions for the integrated branching ratio
$\tau_B\,\langle\Gamma\rangle$, the polarization parameters
$\langle\,f_i\rangle$, and the asymmetries $\langle\,A_{\rm
T}^{(2)}\rangle$, $\langle\,A_{\rm Im}\rangle$ with the
integration boundaries $0.030\,{\rm GeV}\leq m_{ee}\leq 1\,{\rm
GeV} $. The long-distance contribution from $\rho$, $\omega$, and
$\varphi$ mesons is added. FF stands for the form factor model
chosen according to \cite{Ali:2000}, \cite{Ball:2005} and
Eqs.~(\ref{eq:014})--(\ref{eq:016}) and (\ref{eq:0231}). }
\label{tab:smLD1}
\begin{center}
\begin{tabular}{|c|c|c|c|c|c|c|c|}
\hline & &
\multicolumn{3}{|c|}{$m_s=79\,{\rm MeV}$}& \multicolumn{3}{|c|}{$m_s=0$}\\
 \cline{3-8}&$\delta_V$ &FF\cite{Ali:2000} & FF\cite{Ball:2005} &
 FF & FF\cite{Ali:2000}& FF\cite{Ball:2005} & FF \\
 \hline
&$-\pi/4$ &$1.92$&$1.79$&$1.99$&$1.92$&$1.79$&$1.99$\\
$\tau_B\,\langle\Gamma\rangle\times10^{7}$&$0$&$1.92$&$1.79$&$1.99$ &$1.92$ &$1.79$&$1.99$\\
&$\pi/4$ &$1.92$ &$1.79$&$1.99$&$1.91$&$1.79$&$1.99$\\
\hline
&$-\pi/4$ &$0.25$&$0.20$&$0.19$&$0.25$&$0.20$&$0.19$\\
$\langle\,f_0\rangle$&$0$&$0.25$&$0.20$&$0.19$ &$0.25$ &$0.20$&$0.19$\\
&$\pi/4$ &$0.24$&$0.20$&$0.19$&$0.24$&$0.20$&$0.19$\\
\hline
&$-\pi/4$ &$0.39$&$0.41$&$0.42$&$0.38$&$0.40$&$0.40$\\
$\langle\,f_\perp\rangle$ &$0$&$0.39$&$0.41$&$0.42$&$0.38$ &$0.40$&$0.40$\\
&$\pi/4$ &$0.39$&$0.42$&$0.42$ &$0.38$&$0.40$&$0.41$\\
\hline
&$-\pi/4$ &$0.36$&$0.39$&$0.39$&$0.38$&$0.40$&$0.41$ \\
$\langle\,f_\parallel\rangle$  &$0$&$0.36$&$0.39$&$0.39$&$0.38$ &$0.40$&$0.41$\\
&$\pi/4$ &$0.36$&$0.39$&$0.39$&$0.38$&$0.40$&$0.41$\\
\hline
&$-\pi/4$ &$3.5$&$2.9$&$3.6$&$-0.5$&$-1.0$&$-0.4$ \\
$\langle\,A_{\rm T}^{(2)}\rangle\times10^{2}$ &$0$&$3.6$&$3.0$&$3.7$&$-0.4$ &$-1.0$ &$-0.3$\\
&$\pi/4$ &$3.9$&$3.4$&$4.0$&$0.$ &$-0.6$&$0.1$\\
\hline
&$-\pi/4$ &$0.6$&$0.6$&$0.6$&$0.6$&$0.6$&$0.6$ \\
$\langle\,A_{\rm Im}\rangle\times10^{3}$  &$0$ &$-1.2$&$-1.3$&$-1.2$&$-1.2$ &$-1.3$ &$-1.2$\\
&$\pi/4$ &$-2.3$&$-2.4$&$-2.3$&$-2.3$&$-2.4$&$-2.3$\\
\hline
\end{tabular}
\end{center}
\end{table}

\begin{table}[t]
\caption{Same as Table \ref{tab:smLD1} but with the integration
boundaries $0.5\,{\rm GeV}\leq m_{ee}\leq 1\,{\rm GeV} $. }
\label{tab:smLD2}
\begin{center}
\begin{tabular}{|c|c|c|c|c|c|c|c|}
\hline & &
\multicolumn{3}{|c|}{$m_s=79\,{\rm MeV}$}& \multicolumn{3}{|c|}{$m_s=0$}\\
 \cline{3-8}&$\delta_V$ &FF\cite{Ali:2000} & FF\cite{Ball:2005} &
 FF & FF\cite{Ali:2000}& FF\cite{Ball:2005} & FF \\
 \hline
&$-\pi/4$ &$5.7$&$4.8$&$5.3$&$5.7$&$4.8$&$5.2$\\
$\tau_B\,\langle\Gamma\rangle\times10^{8}$&$0$ &$5.6$&$4.8$&$5.2$&$5.6$ &$4.8$ &$5.2$\\
&$\pi/4$ &$5.6$ &$4.7$&$5.2$&$5.6$&$4.7$&$5.2$\\
\hline
&$-\pi/4$ &$0.62$&$0.56$&$0.55$&$0.62$&$0.56$&$0.55$\\
$\langle\,f_0\rangle$&$0$ &$0.62$&$0.56$&$0.54$&$0.62$ &$0.56$&$0.54$\\
&$\pi/4$ &$0.62$&$0.56$&$0.54$&$0.62$&$0.56$&$0.54$\\
\hline
&$-\pi/4$ &$0.20$&$0.23$&$0.24$&$0.19$&$0.22$&$0.23$\\
$\langle\,f_\perp\rangle$ &$0$ &$0.20$&$0.23$&$0.24$&$0.19$ &$0.22$&$0.23$\\
&$\pi/4$ &$0.20$&$0.23$&$0.24$&$0.19$&$0.22$&$0.23$\\
\hline
&$-\pi/4$ &$0.18$&$0.21$&$0.21$&$0.19$&$0.22$&$0.23$ \\
$\langle\,f_\parallel\rangle$  &$0$ &$0.18$&$0.21$&$0.22$&$0.19$ &$0.22$&$0.23$\\
&$\pi/4$ &$0.18$&$0.21$&$0.22$&$0.19$&$0.23$&$0.23$\\
\hline
&$-\pi/4$ &$5.0$&$3.5$&$5.3$&$0.5$&$-1.0$&$0.8$ \\
$\langle\,A_{\rm T}^{(2)}\rangle\times10^{2}$&$0$ &$5.0$&$3.6$&$5.3$&$0.5$ &$-0.9$&$0.9$\\
&$\pi/4$ &$5.0$&$3.6$&$5.4$&$0.6$ &$-0.9$&$0.9$\\
\hline
&$-\pi/4$ &$0.6$&$0.2$&$0.8$&$0.3$&$-0.1$&$0.5$ \\
$\langle\,A_{\rm Im}\rangle\times10^{4}$  &$0$ &$-0.2$&$-0.8$&$-0.1$&$-0.5$ &$-1.1$&$-0.4$\\
&$\pi/4$ &$-0.6$&$-1.1$&$-0.5$&$-0.8$&$-1.5$&$-0.8$\\
\hline
\end{tabular}
\end{center}
 \end{table}

As seen from Tables~\ref{tab:sm1}, \ref{tab:sm2}, the branching
ratio does not depend on the $s$ quark mass, while it is sensitive
to the form factors, especially in region (b).

The value of the $K^*$ polarization fraction $\langle f_0 \rangle$
does not change when varying the mass of the strange quark, while
the polarization fractions $\langle\,f_\perp\rangle, \,
\langle\,f_\parallel\rangle$ show weak dependence on value of
$m_s$. Variations of all fractions with the form factor models are
about $10\%-20\%$. In region (a) the longitudinal polarization is
smaller than the transverse ones, while in region (b) the
longitudinal polarization prevails over the transverse ones.

As for the asymmetry $\langle\,A_{\rm T}^{(2)}\rangle $, one can
notice its strong dependence on the choice of the form factors and
especially on the value of the strange quark mass. Note that for
the form factors, calculated using
Eqs.~(\ref{eq:014})--(\ref{eq:016}) and (\ref{eq:0231}), this
asymmetry is proportional to $m_s$ if one neglects the mass of
$K^*$. Then the $\langle\,A_{\rm T}^{(2)}\rangle $ value in the
last column of these tables would be equal to zero. However, in
our calculation we do not neglect the mass of $K^*$; therefore
$\langle\,A_{\rm T}^{(2)}\rangle \ne 0$ for $m_s =0$. For the
nonzero value of $m_s$ the calculated asymmetry is of the order
3\%--5\% depending on the choice of the form factors. The
asymmetry $\langle\,A_{\rm Im}\rangle $ appears to be very small,
on the level of $10^{-5}$--$10^{-4}$.

Now we discuss the results with the total amplitude, including
resonances (see Tables~\ref{tab:smLD1} and \ref{tab:smLD2}). In
this calculation, for definiteness, the relative resonant phases
$\delta_V$ for $V =\rho, \omega, \phi$ have been taken equal to
each other. For the estimation, we have chosen three values of the
phase: \ $-\pi/4$, \, $0$, and $+\pi/4$.

Let us start with the branching ratio
$\tau_B\,\langle\Gamma\rangle$. As seen by comparing  Tables
\ref{tab:sm1} and \ref{tab:sm2} with Tables \ref{tab:smLD1} and
\ref{tab:smLD2}, in the $q^2$ interval (a) the resonant
contribution is negligibly small. In interval (b) this
contribution is bigger, at the level of 1\%, which is still much
smaller than the expectations of Ref.~\cite{Lefrancois:2009}.

The polarization fractions of the $K^*$ also do not change by more
than $\sim$5\% after inclusion of the resonances, though $\langle
\, f_i \, \rangle$ are more sensitive to the choice of the
transition form factors $B \to K^*$.

On the contrary, the asymmetry $\langle\,A_{\rm T}^{(2)}\rangle$
receives a large contribution from the resonances. In the region
0.030 GeV $<m_{ee}<$ 1 GeV, this contribution can reach up to
$15$\%  depending on the choice of form factors and the resonant
phase $\delta_V$, while in the region 0.5 GeV $<m_{ee}<$ 1 GeV,
the resonant contribution appears to be much smaller, $\sim $3\%.
Of course, the asymmetry remains of the order of a few percent. We
should emphasize the strong dependence of this observable on the
choice of form factors.

This integrated asymmetry remains sensitive to the value of $m_s$,
and therefore sensitive to the wrong helicity transition $b_L \to
s_R + \gamma_R$. It also follows from our calculation that effects
of NP should lead to values of $\langle\,A_{\rm T}^{(2)}\rangle
\gtrsim 0.1$; otherwise, it will be difficult to distinguish these
effects from all model uncertainties discussed above in
Sec.~\ref{subsec:differential}.

As for the asymmetry $\langle\,A_{\rm Im}\rangle$, it changes
drastically after adding the resonances (compare Tables
\ref{tab:sm1} and \ref{tab:sm2} with Tables \ref{tab:smLD1} and
\ref{tab:smLD2}), from values $ \sim 10^{-5}$ without resonances
to values $\sim 10^{-3}$ [in region (a)] and $ \sim 10^{-4}$ [in
region (b)] with resonances. Note that $\langle\,A_{\rm
Im}\rangle$ is determined by the imaginary part of the amplitude.
The latter in the SM (without resonances) is determined by the
light-quark loop through the function $Y(q^2)$ \cite{BFS:2001},
and therefore the imaginary part of the nonresonant amplitude
appears to be very small, $\sim 10^{-5}$. It is not surprising
that the imaginary part of the total amplitude in
Eq.~(\ref{eq:032}) is determined solely by the resonant
contribution.

Of course, this observable strongly depends on the resonant phase
$\delta_V$; however, for any phase it remains small. Since this
asymmetry is determined mainly by the resonant amplitude, it does
not show prominent dependence on $m_s$, especially in region (a).
For these reasons $\langle\,A_{\rm Im}\rangle$ is not very
suitable for the study of the chiral structure of the decay
amplitude. At the same time, the calculation shows that
observation of this asymmetry at the level of $\sim 1$\% or bigger
will indicate effects beyond the SM.


\section{ \label{sec:conclusions}
 Conclusions}

Branching ratios and other observables for the rare FCNC decay
${\bar B}_d^0 \to {\bar K}^{*0} \, (\to K^{-}\, \pi^+) \, e^+\,e^-
$ have been studied in the region of electron-positron invariant
mass below the $\bar{c}c$ threshold. Our main emphasis has been
placed on an accurate account of the mechanism $\bar{B}_d^0 \to
\bar{K}^{*0} \, (\to K^{-}\, \pi^+) \, V $ with low-lying vector
resonances $V = \rho(770), \, \omega(782), \, \phi(1020)$ decaying
into the $e^+ e^-$ pair.

The invariant-mass dependence of the branching ratio and
coefficients in the angular distribution of the lepton pair,
$A_{\rm T}^{(2)}$, \ $A_{\rm Im}$, \ $d \bar{A}_{\rm FB}/d q^2$,
has been calculated and studied. In view of the planned
experiments at the LHCb, in which the observables integrated over
the invariant mass  will be measured~\cite{Lefrancois:2009}, we
also calculated the corresponding quantities.

In general, the resonant contribution appears to be small in the
branching ratio, polarization parameters of the $K^*$ meson, and
forward-backward asymmetry. Nevertheless, some of the observables
change drastically after adding the resonances to the pure SM
contribution. In particular, the $q^2$ dependence of the asymmetry
$A_{\rm T}^{(2)}$ gets considerably modified by the vector
resonances. This observation is of importance in view of the
sensitivity of $A_{\rm T}^{(2)}$ to the value of the strange quark
mass, and thereby to the chiral-odd dipole transition $b_L \to s_R
+ \gamma_R$. Thus  $A_{\rm T}^{(2)}$ is also sensitive to effects
of NP which are related to the right-handed currents. Still,
$A_{\rm T}^{(2)}$ in the SM with resonances is small, of the order
of a few percent. The resonances also increase the asymmetry
$A_{\rm Im}$ by at least $1$ order of magnitude; however, this
observable remains very small, $10^{-4}-10^{-3}$, and therefore it
is difficult to measure.

There is dependence of the calculated quantities on the model of
transition form factors which have been considered. In general,
the corridor due to different models is of the order of $\sim
$5\%. In this connection, we have introduced a new
forward-backward asymmetry $d \widetilde{A}_{\rm FB}/d q^2$,
normalized differently compared to the standard definition. This
modified forward-backward asymmetry has the advantages of being
almost independent of the form factor model and of taking big
values up to $\pm 0.75$.

Most of the above features remain after integration of the observables over the
$e^+ e^-$ invariant mass up to 1~GeV. Two integration regions have been
selected which are particularly suitable for the planned future measurements at
the LHCb~\cite{Lefrancois:2009}. The predictions for all integrated observables
are given in the framework of the SM, taking into account of low-lying vector
resonances. \vspace{0.7cm}

\appendix

\section{\label{sec:Appendix} Matrix element and form factors}

\subsection{\label{subsec:matrix element} Matrix element}

The effective Hamiltonian for the quark-level transition $ b \to
s\, e^+ e^- $ within the SM is well-known and can be taken, e.g.,
from Ref.~\cite{Antonelli:2009}. It is expressed in terms of the
local operators ${\cal O}_i$ and Wilson coefficients $C_i$, where
$i=1, \ \ldots, \ 6, \ 7\gamma,\ 8g,\ 9V, 10A$.

The matrix element of this effective Hamiltonian for the
nonresonant decay ${\bar B}_d^0 (p)\to {\bar
K}^{*0}(k,\epsilon)\,e^+(q_+)\,e^-(q_-)$ can be written, in the
so-called naive factorization~\cite{Antonelli:2009}, as
\begin{widetext}
\[
{\cal M}_{\rm NR} = \frac{G_F\alpha_{\rm
em}}{\sqrt{2}\pi}\,V_{ts}^* V_{tb}\Bigl(\langle{\bar
K}^{*0}(k,\epsilon)|\bar{s} \gamma_\mu P_L b|{\bar
B}_d^0(p)\rangle\bigl(C_{9V}^{\rm eff} \bar{u}(q_-) \gamma^\mu
v(q_+) +C_{10A}\bar{u}(q_-) \gamma^\mu \gamma_5 v(q_+)\bigr)\]
\begin{equation}\label{eq:005}
-\frac{2}{q^2}C_{7\gamma}^{\rm eff}\langle{\bar
K}^{*0}(k,\epsilon)|\bar{s}\,i\, \sigma_{\mu\nu}
q^\nu(\overline{m}_b(\mu)P_R +\overline{m}_s(\mu)P_L)\, b|{\bar
B}_d^0(p)\rangle\,\bar{u}(q_-) \gamma^\mu v(q_+)\Bigr)\,.
\end{equation}
\end{widetext}

Here, $V_{ij}$ are the Cabibbo-Kobayashi-Maskawa matrix elements
\cite{CKM}, $G_F$ is the Fermi coupling constant, $\alpha_{\rm
em}$ is the electromagnetic fine-structure constant,
$P_{L,R}=(1\mp \gamma_5)/2$ denote chiral projectors, and
$\overline{m}_b(\mu)$ [$\overline{m}_s(\mu)$] is the running
bottom (strange) quark mass in the $\overline{\rm MS}$ scheme at
the scale $\mu$. Moreover, $\sigma_{\mu \nu}={i\over
2}[\gamma_{\mu},\gamma_{\nu}]$, $q_{\mu}=(q_{+} +q_{-})_{\mu}$,
$C_{7\gamma}^{\rm eff} = C_{7\gamma}-(4 \bar C_3-\bar C_5)/9-(4
\bar C_4-\bar C_6)/3$, $C_{9V}^{\rm eff} = C_{9V} +Y(q^2)$, where
$Y(q^2)$ is given in Ref.~\cite{BFS:2001}.

The ``barred'' coefficients $\bar{C}_i$ (for $i=1,\ldots,6$) are
defined as certain linear combinations of the $C_i$, such that the
$\bar{C}_i$ coincide at leading logarithmic order with the Wilson
coefficients in the standard basis \cite{Buchalla:1996}. The
coefficients $C_i$ are calculated at the scale $\mu=m_W$, in a
perturbative expansion in powers of $\alpha_s(m_W)$, and are then
evolved down to scales $\mu \sim m_b$ using the renormalization
group equations.

The $\overline{\rm MS}$ mass $\overline{m}_b(\mu)$ can be related
with the pole mass $m_b$ at the scale $\mu=m_b$ through
\cite{Gray:1990, Chetyrkin:1999}
\[\overline{m}_b(m_b)=m_b\Bigl(1-{4\over 3}{\alpha_s(m_b)\over \pi}
-10.167\,\Bigl({\alpha_s(m_b)\over \pi}\Bigr)^2 \]
\[+{\cal
O}\Bigl(\Bigl({\alpha_s(m_b)\over \pi}\Bigr)^3\Bigr) \Bigr)\,.\]
The expression for the next terms in this equation can be found in
Ref.~\cite{Chetyrkin:1999}. The mass of the strange quark can be
determined from the spectral function sum rules or lattice QCD
simulation \cite{Manohar:PDG}. The up-to-date value of $m_s$ given
by the PDG \cite{PDG:2008} is $\overline{m}_s(2\,{\rm GeV})=95\pm
25 \,{\rm MeV}$. Note that this running mass is evaluated at
$\mu_0=2\,{\rm GeV}$ with three active quark flavors. The
evolution of the $\overline{m}_s(\mu)$ is governed by the
renormalization group equation which has the solution
\cite{Chetyrkin:1997}
\[\frac{\overline{m}_s(\mu)}{\overline{m}_s(\mu_0)}=
\frac{f(\alpha_s(\mu)/\pi)}{f(\alpha_s(\mu_0)/\pi)}\,,\] with
\[f(x)={\displaystyle x^{4\over 9}}(1+0.895062\,x+1.37143\,x^2 \,
+ {\cal O}(x^3))\,.\]

\subsection{\label{subsec:form factors}
Form factors of $B \to K^*$ transition}

The hadronic part of the matrix element in Eq.~(\ref{eq:005})
describing the $B\to K^*e^+e^-$ transition can be parametrized in
terms of $B\to K^*$ form factors, which usually are defined as
\begin{equation}\label{eq:009}
\langle\bar{K}^*(k,\epsilon)|\bar{s} \gamma_\mu
b|\bar{B}(p)\rangle=\frac{2\,V(q^2)}{m_B+m_{K^*}}\,
\varepsilon_{\mu\nu\alpha\beta}\,\epsilon^{\nu*}p^{\alpha}k^{\beta}\,,
\end{equation}
\[\langle\bar{K}^*(k,\epsilon)|\bar{s} \gamma_\mu\gamma_5
b|\bar{B}(p)\rangle=i\epsilon_\mu^*(m_B+m_{K^*})A_1(q^2) \]
\[-i (\epsilon^*\cdot p)(p+k)_\mu\frac{A_2(q^2)}{m_B+m_{K^*}}\]
\begin{equation}\label{eq:010}
-i(\epsilon^*\cdot
p)\,q_\mu\frac{2\,m_{K^*}}{q^2}(A_3(q^2)-A_0(q^2)) \,,
\end{equation}
with
\[A_3(q^2)=\frac{m_B+m_{K^*}}{2\,m_{K^*}}A_1(q^2)-
\frac{m_B-m_{K^*}}{2\,m_{K^*}}A_2(q^2)\,,\]
\[A_0(0)=A_3(0)\,;\]
\begin{equation}\label{eq:011}
\langle\bar{K}^*(k,\epsilon)|\bar{s}\,\sigma_{\mu\nu} q^\nu
b|\bar{B}(p)\rangle=i\,2\,T_1(q^2)\,
\varepsilon_{\mu\nu\alpha\beta}\,\epsilon^{\nu*}p^{\alpha}k^{\beta}\,,
\end{equation}
\[\langle\bar{K}^*(k,\epsilon)|\bar{s}\,\sigma_{\mu\nu}\gamma_5
q^\nu b|\bar{B}(p)\rangle=T_2(q^2)(\epsilon_\mu^*(P\cdot q)\]
\begin{equation}\label{eq:012}
-(\epsilon^*\cdot q)P_\mu)+T_3(q^2)(\epsilon^*\cdot
q)(q_\mu-\frac{q^2}{P\cdot q}P_\mu)\,,
\end{equation}
with $T_1(0)=T_2(0)$. In the above equations, $q=p-k$, $P=p+k$,
$p^2=m_B^2$, $k^2=m_{K^*}^2$, $\epsilon^\mu$ is the polarization
vector of the $K^*$ meson, $\epsilon^*\cdot k=0$, and
$\varepsilon_{0123}=1$.

\begin{table}[t]
\caption{Input values for the parametrization
  (\ref{eq:013}) of the $B\to K^*$ form factors.}
\label{tab:ff1}
\begin{center}
\begin{tabular}{c c c c c c c c}
\hline \hline & $A_1$ &$ A_2$ &$A_0$ & $V$ & $T_1$ & $T_2$ & $T_3$\\
\hline
$F(0)$ & $0.294$ & $0.246$ & $0.412$ &$0.399$ & $0.334$ & $0.334$ & $0.234$\\

$c_1$ & $0.656$ & $1.237$ & $1.543$ & $1.537$ & $1.575$ & $0.562$ & $1.230$\\

$c_2$ & $0.456$ & $0.822$ & $0.954$ & $1.123$ & $1.140$ & $0.481$ & $1.089$\\
\hline \hline
\end{tabular}
\end{center}
\end{table}
In the numerical estimations, we use the form factors from LCSR
calculations \cite{Ali:2000} and \cite{Ball:2005} as well as the
large-energy-effective-theory form factors $\xi_\perp(q^2)$ and
$\xi_\parallel(q^2)$ \cite{Charles:1999, BFS:2001, Beneke:2001}.
Form factors given in \cite{Ali:2000} are parametrized as follows
\begin{equation}\label{eq:013}
F({q}^2)=F(0)\exp(c_1\hat{q}^2+c_2\hat{q}^4)\,,
\end{equation}
where $\hat{q}^2\equiv q^2/m_B^2$. The coefficients in this
parametrization are listed in Table \ref{tab:ff1}. The $q^2$
dependence of the $B\to K^*$ form factors given in
\cite{Ball:2005} is parametrized as
\begin{equation}\label{eq:014}
F(q^2)=\frac{r_1}{1-q^2/m_R^2}+\frac{r_2}{1-q^2/m_{fit}^2}\,,
\end{equation}
\begin{equation}\label{eq:015}
F(q^2)=\frac{r_1}{1-q^2/m_{fit}^2}+\frac{r_2}{(1-q^2/m_{fit}^2)^2}\,,
\end{equation}
\begin{equation}\label{eq:016}
F(q^2)=\frac{r_2}{1-q^2/m_{fit}^2}\,,
\end{equation}
where the fit parameters $r_{1,2}$, $m_R^2$, and $m_{fit}^2$ are
shown in Table \ref{tab:ff2}.
\begin{table}[t]
\caption{The parameters $r_{1,2}$, $m_R^2$, and $m_{fit}^2$
describing the $q^2$ dependence of the $B\to K^*$ form factors in
the LCSR approach \cite{Ball:2005} and
$T_3(q^2)=\displaystyle\frac{m_B^2-m_{K^*}^2}{q^2}\left(\widetilde{T}_3(q^2)-
T_2(q^2)\right) $. The fit equations to be used are given in the
last column. } \label{tab:ff2}
\begin{center}
\begin{tabular}{c c c c c c}
\hline \hline & $r_1$ &$ r_2$ &$m_R^2\,,\rm GeV^2$ & $m_{fit}^2\,,\rm GeV^2$ & Fit eq.\\
\hline
$V$   & $0.923$ & $-0.511$ & $(5.32)^2$ &$49.40$ & (\ref{eq:014})\\

$A_1$ &         & $0.290$ &             & $40.38$ &(\ref{eq:016})\\

$A_2$ &$-0.084$ & $0.342$ &             & $52.00$ &(\ref{eq:015})\\

$A_0$ & $1.364$ & $-0.990$ & $(5.28)^2$ &$36.78$ & (\ref{eq:014})\\

$T_1$ & $0.823$ & $-0.491$ & $(5.32)^2$ &$46.31$ & (\ref{eq:014})\\

$T_2$ &         & $0.333$ &             & $41.41$ &(\ref{eq:016})\\

$\widetilde{T}_3$ &$-0.036$ & $0.368$ &           & $48.10$&(\ref{eq:015})\\
\hline \hline
\end{tabular}
\end{center}
 \end{table}
In the large-energy effective theory the seven \emph{a priori}
independent $B\to K^*$ form factors in
Eqs.~(\ref{eq:009})--(\ref{eq:012}) can be expressed in terms of
two universal form factors $\xi_\perp(q^2)$ and
$\xi_\parallel(q^2)$ \cite{Charles:1999}:
\begin{equation}\label{eq:017}
 A_1(q^2)= \frac{2 E_{K^*}}{m_B + m_{K^*}}\,\xi_{\perp}(q^2)\, ,
\end{equation}
\begin{equation}\label{eq:018}
A_2(q^2)= \frac{m_B+m_{K^*}}{m_B} \left(\xi_{\perp}(q^2)-
\xi_\parallel(q^2)\right)\, ,
\end{equation}
\begin{equation}\label{eq:019}
A_0(q^2) =\frac{E_{K^*}}{m_{K^*}}\,\xi_\parallel(q^2)+
\frac{m_{K^*}}{m_B}\left(\xi_\perp(q^2)-\xi_\parallel(q^2)\right)\,
,
\end{equation}
\begin{equation}\label{eq:020}
V(q^2)= \frac{m_B+m_{K^*}}{m_B}\,\xi_\perp(q^2)\, ,
\end{equation}
\begin{equation}\label{eq:021}
T_1(q^2)=\xi_\perp(q^2)\, ,
\end{equation}
\begin{equation}\label{eq:022}
T_2(q^2)=\left(1-\frac{q^2}{m_B^2-m_{K^*}^2}\right)\xi_\perp(q^2)\,,
\end{equation}
\begin{equation}\label{eq:023}
T_3(q^2)=\xi_\perp(q^2) - \left(1-\frac{m_{K^*}^2}{m_B^2}\right)
\xi_\parallel(q^2)\,.
\end{equation}
Note the different convention for the longitudinal form factor
with $\xi_\parallel(q^2)=m_{K^*}/E_{K^*}\zeta_\parallel(q^2)$,
$\zeta_\parallel(q^2)$ being defined in Ref.~\cite{Charles:1999}.
Here, $E_{K^*}$ is the energy of the final vector meson in the
${B}$ rest  frame,
\[
E_{K^*}=\frac{m_B}{2}\left(1-\frac{q^2}{m_B^2}+
\frac{m_{K^*}^2}{m_B^2} \right).
\]
The form factors $\xi_\perp(q^2)$ and $\xi_\parallel(q^2)$ are
defined by the relations
\[\xi_\perp(q^2)= \frac{m_B}{m_B+m_{K^*}}\,V(q^2)\,,\]
\begin{equation}\label{eq:0231}
\xi_\parallel(q^2)= \frac{m_B + m_{K^*}}{2
E_{K^*}}\,A_1(q^2)-\frac{m_B}{m_B+m_{K^*}}\,A_2(q^2)\, .
\end{equation}
We use the definitions Eqs.~(\ref{eq:014})--(\ref{eq:016}) and
(\ref{eq:0231}), with parameters given in Table \ref{tab:ff2}, to
determine the $q^2$ dependence of $\xi_\perp $ and
$\xi_\parallel$.


\end{document}